# Model for l/f Flux Noise in SQUIDs and Qubits

Roger H. Koch,[1] David P. DiVincenzo[1] and John Clarke[2]

[1]*IBM Research Division, Thomas J. Watson Research Center, Yorktown Heights, NY 10598*

[2]*Department of Physics, University of California, Berkeley, CA 94720-7300 and*

*Materials Sciences Division, Lawrence Berkeley National Laboratory, Berkeley CA 94720*

(Dated: May 5, 2007)

We propose a model for 1/f flux noise in superconducting devices (f is frequency). The noise is generated by the magnetic moments of electrons in defect states which they occupy for a wide distribution of times before escaping. A trapped electron occupies one of the two Kramers-degenerate ground states, between which the transition rate is negligible at low temperature. As a result, the magnetic moment orientation is locked. Simulations of the noise produced by randomly oriented defects with a density of $5 \times 10^{17}$ m$^{-2}$ yield 1/f noise magnitudes in good agreement with experiments.

The phenomenon of 1/f noise, with spectral density $S(f)$ scaling inversely with frequency $f$, is common to virtually all devices. In 1983, Koch et al. [1] identified two separate sources of 1/$f$ noise in dc SQUIDs (Superconducting QUantum Interference Devices): critical current noise and flux noise. The 1/$f$ flux noise $S_\Phi^{1/2}$ (1 Hz) was within a factor of 3 of 10 $\mu\Phi_0$ Hz$^{-1/2}$ for Nb- or Pb-based SQUIDs at 4.2 K, even though the loop areas ranged over 6 orders of magnitude; here, $\Phi$ denotes magnetic flux and $\Phi_0 \equiv h/2e$ is the flux quantum. Subsequently, other authors found rather lower levels of 1/$f$ flux noise at 1 Hz and 4.2 K, for example, 0.5 $\mu\Phi_0$ Hz$^{-1/2}$ [2] and 0.2 $\mu\Phi_0$ Hz$^{-1/2}$ [3]. Wellstood et al. [4] reported values of $S_\Phi^{1/2}$ (1 Hz) of (4-10) $\mu\Phi_0$ Hz$^{-1/2}$ at temperatures below 0.1 K in 12 Nb, Pb and PbIn devices. Recently, Yoshihara et al. [5] showed that 1/f flux noise with $S_\Phi^{1/2}$ (1 Hz) ≈ 1 $\mu\Phi_0$ Hz$^{-1/2}$ determined the decoherence time in their Al-based flux qubits at 20 mK. The value of $S_\Phi^{1/2}$ (1 Hz) in the SQUIDs of Wellstood et al., with areas up to 2 x 10$^5$ $\mu$m$^2$, is at most one order of magnitude higher than that in these qubits, with an area of about 3 $\mu$m$^2$, five orders of magnitude less. These results, and that of Ref. 1, rule out the notion of a "global magnetic field noise".

Critical current fluctuations in Josephson junctions have been widely studied, for example [6-8], and are understood to arise from the trapping and release of electrons in traps in the tunnel barrier. In the case of high transition temperatures ($T_c$) SQUIDs at 77 K, 1/f flux noise is ascribed to thermal activation of vortices among pinning sites [9]. This noise can be eliminated by reducing the linewidth to below $(\Phi_0/B)^{1/2}$, thereby making it energetically unfavorable for the film to trap a vortex [10]; $B$ is the magnetic field in which the device is cooled. Given that the low-$T_c$ devices are made of films with a much higher pinning energy, are operated at much lower temperatures, and may have linewidths orders of magnitude less than $(\Phi_0/B)^{1/2}$, vortex motion is not a viable mechanism for their 1/$f$ flux noise. Thus, the origin of 1/$f$ flux noise in low-$T_c$ devices–despite its ubiquitous nature and the limitations it imposes on SQUIDs and qubits alike–has remained an unsolved puzzle.

In this paper, we propose a model for 1/$f$ flux noise in low-$T_c$ devices. Our basic assumption is that the noise is generated by unpaired electrons that hop on and off defect centers by thermal activation. The spin of an electron is locked in direction while the electron occupies a given trap; this direction varies randomly from trap to trap. The relevant trapping energies have a broad distribution on the scale of $k_B T$ [11], so that the characteristic times over which an electron resides on any one defect vary over many orders of magnitude. The uncorrelated changes of these spin directions yield a series of random telegraph signals that sum to a 1/$f$ power spectrum [12]. There is no shortage of candidates for microscopic defect centers involved in this process: In amorphous $SiO_2$, these include E′ center variants, in which an electron is captured by a silicon atom that has an oxygen vacancy, the nonbridging oxygen hole center (NBOHC) where a hole is trapped on an oxygen atom that has only one bond with the lattice, and the superoxide radical, in which a hole is trapped on an additional oxygen atom [13]. In addition, although not nearly as extensively studied as $SiO_2$, the amorphous oxides of superconductors such as $AlO_x$ and $NbO_x$ contain large densities of defects of various sorts: for example, the concentration of OH defects in $AlO_x$ can reach several percent [14, 15].

Elucidation of this model involves two key steps. First one has to understand how the direction of an electron spin can remain fixed for very long periods of time–longer than the inverse of the lowest frequency at which the 1/$f$ noise is observed, say, $10^{-4}$ Hz. Second, one has to calculate the net fluctuating flux coupled into a superconducting loop. We address these two issues in turn.

Our key assumption is that an electron randomly adopts a low-energy spin direction when it arrives at a defect, and that it remains *locked* in that orientation during its entire residence time. If the magnetic field ***B*** is zero, Kramers' theorem [16] guarantees that the ground state is doubly degenerate, the two states having oppositely directed angular momenta [Fig. 1(a)]. It is well known that scattering mechanisms that take the electron from one member of the doublet to the other are extremely weak: the "Van Vleck cancellation" [17] implies that direct phonon scattering is forbidden. Higher order processes are allowed, but those that have been studied are strongly suppressed at low temperature; for example, the phonon Raman scattering rate [18] has a temperature dependence of $T^{13}$

Of course, the magnetic field is not strictly zero; any particular defect experiences fluctuating dipole fields from neighboring defects of the order of $10^{-4}$ T (root mean square). The magnetic moment vector of the defect $\hat{\boldsymbol{M}} = \mu_B (\hat{\boldsymbol{L}} + 2\hat{\boldsymbol{S}})$ can be locked as a result of spin-orbit coupling, making it stable with respect to these field fluctuations. The following model Hamiltonian [19] provides a good generic description of this locking effect:

$$\hat{H} = \sum_{i=x,y,z} V_i |p_i\rangle\langle p_i| + \lambda \hat{\boldsymbol{L}} \cdot \hat{\boldsymbol{S}} + \mu_B \boldsymbol{B} \cdot (\hat{\boldsymbol{L}} + 2\hat{\boldsymbol{S}}). \qquad (1)$$

In this model, the unpaired electron occupies a p-orbital; the $V_{x,y,z}$ are the matrix elements of the crystal field potential (there will be a preferred coordinate system, varying randomly from defect to defect, for which the crystal-field tensor is diagonal, as shown). The spin-orbit coupling constant $\lambda$ is observed to have a large range of possible magnitudes for different defects, in the range of [19] 10 K to 5000 K, but for defects involving atomic weights near that of silicon, $|\lambda| \approx 300 K$ is typical [19]. The scale of the crystal field parameters $V_{x,y,z}$ is set by chemical energies, and so can range up to [19] $\approx 2000K$. It is often said that the orbital angular momentum of simple defects is "quenched" [20], meaning that $<\hat{L}> = 0$ and that the magnetic moment arises only from the (unlocked) spin angular momentum. Equation (1) exhibits this behavior if $|V_i - V_j| >> |\lambda|$ ($i \neq j$ = x,y,z). But, it seems quite reasonable that there is a substantial subpopulation of defects for which $|V_i - V_j| \approx |\lambda|$, and for these Figs. 1(b) and (c) show that the direction of $\boldsymbol{M} = \langle \Psi_0 | \hat{M} | \Psi_0 \rangle$ for the ground state $|\Psi_0\rangle$ is very stable with respect to variations in the direction of a $10^{-4}$-T magnetic field, being locked to the principal axis of the crystal field. In defects for which $\lambda < 0$, $\boldsymbol{L}$ and $\boldsymbol{S}$ are parallel and $\boldsymbol{M}$ is large, while for the $\lambda > 0$ defects, $\boldsymbol{M}$ is near zero (i.e., the anisotropic Lande g-factor is near zero) because $\boldsymbol{L}$ and $\boldsymbol{S}$ are antiparallel; thus, we expect the $\lambda < 0$ subpopulation to be most important for flux noise.

Given this picture of the underlying physical processes, we now calculate the flux noise coupled into a SQUID or qubit (henceforth succinctly referred to as "SQUID") by a spatially random distribution of electron spins fluctuating in orientation. We assume–for lack of more specific information–that the defects are randomly distributed over the substrate, everywhere with the same areal density $n$. We consider three regions that produce noise (inset, Fig. 2): the hole of the SQUID ("hole noise"), the region outside the SQUID ("exterior noise") and the loop itself ("loop noise") [21]. For purposes of simulating the coupling between an electron magnetic moment and the SQUID, we represent the moment by a small test current loop. The SQUID loop lies in the plane z = 1μm, has inner and outer dimensions of 2d and 2D, and a thickness of 0.1μm. We consider the current loop in the plane z = 0 ("perpendicular (p) moment") or in the x- or y-plane, centered at z = 0 ("inplane

(i) moment"). The test loop has an effective area $A = h^2 = (0.1~\mu m)^2$, a strip width $s = 0.03$ µm, a thickness of 0.1µm, and carries a current $i$ chosen so that $Ai = \mu_B$, where $\mu_B = 9.27 \times 10^{-24}~JT^{-1}$ is the Bohr magneton (the scale of the magnetic moment of the defects modeled above). For the test loop at a specified location, we compute its mutual inductances $M_p$ and $M_i$ with the SQUID loop using the superconducting version of FastHenry [22]. The flux coupled into the SQUID for a single electron moment is given by $\Phi_s = M(x,y)i = M(x,y)\mu_B/A$. We calculate the quantity $\Phi_s/\mu_B = M(x,y)/A$—the flux per Bohr magneton coupled into the SQUID loop.

In Fig. 2 we plot $\Phi_s(x,y)/\mu_B$ versus $x$ for $y = 0$ for the magnetic moment perpendicular to the plane and inplane. As expected, the plots are symmetric about the origin. For the perpendicular moment, $\Phi_s(x,y)/\mu_B$ has a local minimum at the center, and increases towards either edge of the superconductor. When the moment is at the midpoint under (or over) the superconducting film, the coupled flux is essentially zero as expected from symmetry. The flux coupled into the SQUID loop from an exterior moment also peaks at the edges of the superconductor. For the inplane moment, $\Phi_s(x,y)/\mu_B$ peaks at the midpoints of the superconducting film, and falls off rapidly as the moment moves away from the film. By symmetry, away from the superconducting region the flux would be zero if the moment and the SQUID loop were in the same plane. In the course of these simulations, we showed that the results did not change when the area of the current loop was varied between $0.1~A$ and $10~A$.

To obtain the noise due to an ensemble of spins, we first integrate $M_p$ and $M_i$ over an element $dxdy$ in one quadrant. We cut off the integration at a distance $L = 100$ µm beyond the outer edge of the SQUID, where $M_p$ or $M_i$ is two orders of magnitude less than at (0,0).

For either case, the total mean square normalized flux noise coupled into the SQUID, summed over the hole, superconductor and exterior contributions, is given by

$$\langle(\delta\Phi_s)^2\rangle = 8n\mu_B^2 \int_0^{(L+D)} dx \int_0^x dy\, [M(x,y)/A]^2. \quad (2)$$

The total mean square noise is given by $\langle(\delta\Phi_{st})^2\rangle = [\langle(\delta\Phi_{sp})^2\rangle + \langle(\delta\Phi_{si,x})^2\rangle + \langle(\delta\Phi_{si,y})^2\rangle]/3$, the angular average of the quadrature sum of the noise from the three coordinate directions. To convert $\langle(\delta\Phi_{st})^2\rangle$ to a spectral density $S_\Phi(f) = \alpha/f$, where $\alpha$ is a constant, we introduce lower and upper cut-off frequencies, $f_1$ and $f_2$, and set $\langle(\delta\Phi_{st})^2\rangle = \alpha\int_{f_1}^{f_2} df/f = \alpha\ln(f_2/f_1)$. Taking $f_1 = 10^{-4}$ Hz and $f_2 = 10^9$ Hz (the results are only weakly sensitive to these values), we find $S_\Phi(f)/\Phi_0^2 \approx \langle(\delta\Phi_s/\Phi_0)^2\rangle/30f$.

We obtain noise levels in reasonable agreement with observations for $n = 5 \times 10^{17}$ m$^{-2}$. This value is six orders of magnitude higher than the value of about $10^{12}$ m$^{-2}$ reported from measurements of two level systems in Josephson junctions [14]. However, the two situations are physically very different. The SiO$_2$ layer on a Si wafer is typically 100 nm thick and, because of its exposure to processing chemicals and the atmosphere, is covered with contaminants that are likely to be highly disordered. In contrast, the thickness of the tunnel barrier is 2-3 nm, and the barrier is protected with a metallic layer immediately after its formation, before it is exposed to any contaminants. Whereas for a 2-nm thickness, an areal density of $5 \times 10^{17}$ m$^{-2}$ corresponds to 1 defect in 300 atoms, for a 100-nm thickness it corresponds to 1 defect in $10^4$ atoms, which does not seem unreasonable for a thick, highly disordered and contaminated material. This areal density is also comparable with estimates of trap densities on silicon surfaces that have been exposed to atmospheric-like conditions [23]. Furthermore, room temperature scanning tunneling microscope experiments [24] on ultraclean silicon surfaces that were exposed to a low level of oxygen in an ultrahigh vacuum system revealed as many as eight near-surface two-level systems in an area of $4 \times 10^{-17}$ m$^2$ (i.e., a density of $2 \times 10^{17}$ m$^{-2}$) in a 10-500 Hz bandwidth, corresponding to $= 2 \times 10^{18}$ m$^{-2}$ over 13 decades of frequency. Thus, although our required value of $n$ awaits experimental verification, *a priori* it does not seem beyond the realm of possibility.

We plot the normalized amplitude spectra of the flux noise in the SQUID at 1 Hz, $S_\Phi^{1/2}$ (1 Hz)/$\Phi_0$, in Fig. 3. Figure 3(a) shows the contributions of the hole, loop and exterior noises for the perpendicular moments, the loop noise for the inplane moments and the total noise versus the mean loop size $D + d$ for constant aspect ratio $2d/W$. All the contributions follow the same general trend, increasing by a factor of 4 when the loop area is increased by a factor of about 200. Figure 3(b) shows the same noise contributions versus $(D + d)$ for fixed $W$. As expected, the hole noise vanishes as the area of the hole vanishes. At values of $(D + d)$ greater than about 50 µm, the slope tends asymptotically to 0.5. This result implies that $S_\Phi$(1 Hz) scales with the linear dimension of the SQUID, that is, with the perimeter rather than the area. Thus, once the dimensions of the hole exceed the strip width, the noise is dominated by defects relatively close to or underneath (or on top of) the superconductor, and the contributions from the central region of the loop become unimportant. The total noise ranges from about 0.7 to 2.5 µ$\Phi_0$ Hz$^{-1/2}$ over the range shown.

Figure 3 (c) and (d) show the dependence of the 1/f noise generated by the perpendicular moments and one direction of inplane moments versus the separation $z_0$ between the current and SQUID loops. For the cases of perpendicular moments and the inplane moments under the superconductor, the noise is independent of $z_0$ for values below about 3 µm. For the hole and exterior inplane moments, as expected the noise drops off as $z_0$ tends to zero; we neglect these contributions in calculating the total noise in Figs. 3(a) and (b). Thus, our choice of $z_0 = 1$ μm in our simulations is well justified.

We briefly discuss the possibility that a similar model based on fluctuating *nuclear* spins in the superconductor or substrate could explain flux noise; we emphasize, however, that we do not have a model in which nuclear fluctuations produce a 1/f power spectrum. As an example, we consider a 100-nm thick film of $^{27}$Al (6 x $10^{28}$ nuclei / m$^3$, magnetic moment 0.0020 $\mu_B$). If we place all these nuclear spins in a 100-nm thick layer under the film (overestimating their effect), the spectral density of the noise would be a factor of 20 lower than for our assumed areal density of electrons. Furthermore, Wellstood *et al.* measured the 1/f noise in SQUIDs with loops made of both Nb (5.6 x $10^{28}$ nuclei / m$^3$, magnetic moment 0.0034 $\mu_B$) and $^{207}$Pb (3.3 x $10^{28}$ nuclei / m$^3$, abundance 22%, magnetic moment 0.00032 $\mu_B$), and found that the noise powers at 0.1 K differed by no more than a factor of 4. Scaling the parameters in our model predicts that the noise power for Nb would be higher than for Pb by a factor of 850. As an example for the substrate, we consider $^{29}$Si (5 x $10^{28}$ nuclei / m$^3$, abundance 5%, magnetic moment 0.00030 $\mu_B$). Taking the results from the example in Fig. 3(c, d), we assume that the nuclei contribute to a depth of 10 μm. The resulting noise power is lower than that for electrons by a factor of 200. We conclude that nuclei are unlikely to be contributors to 1/f flux noise.

In summary, we have shown that flux noise in superconducting devices can be explained in terms of electrons that hop between traps in which their spins have fixed, random orientations. The crucial underlying physics of "locking" is that the ground state of the defect is two-fold degenerate–the Kramers' degeneracy–and that transitions between these states do not occur at low temperature. The assumptions that the traps have a broad spectrum of energies, resulting in a wide range of characteristic trapping times, and that the processes are uncorrelated, lead to 1/$f$ flux noise. The fact that the noise amplitude scales only weakly with area–as the fourth root in the limit where the hole dimension is greater than the strip width–is consistent with experimental observations. The computed 1/f noise magnitude requires a trap density of $5 \times 10^{17}$ m$^{-2}$ to agree with experimental values. It is noteworthy that the two SQUIDs with the lowest 1/$f$ noise [2,3] were passivated. Our picture unifies the concepts of charge, critical current and flux noise: All three noise sources originate in the random filling and emptying of electron traps; flux noise, in addition, involves the concept of spin locking and the random direction of the magnetic moment associated with the trapped electron or hole.

We are grateful to Matt Copel, Jim Hannon and Sohrab Ismail-Beigi for helpful discussions. DDV is supported by the DTO through ARO contract number W911NF-04-C-0098; JC is supported by the Director, Office of Science, Office of Basic Energy Sciences, Materials Sciences and Engineering Division, of the U.S. Department of Energy under Contract No. DE-AC02-05CH11231.

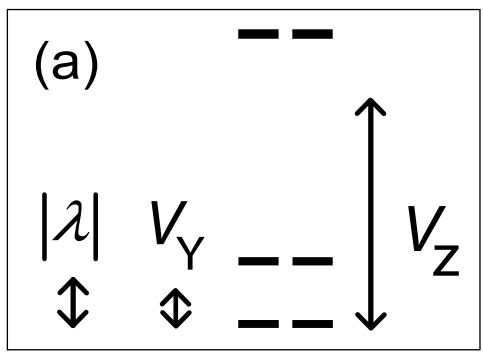
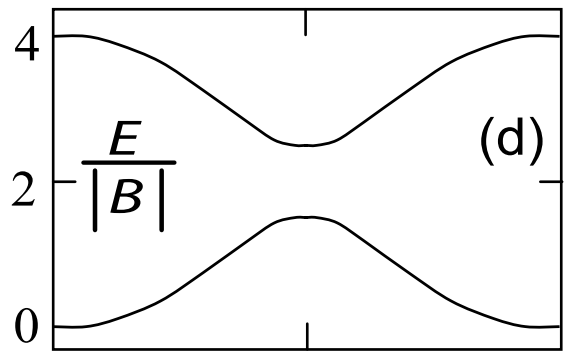
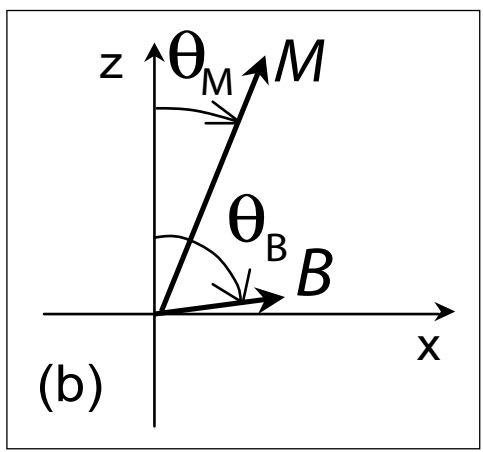
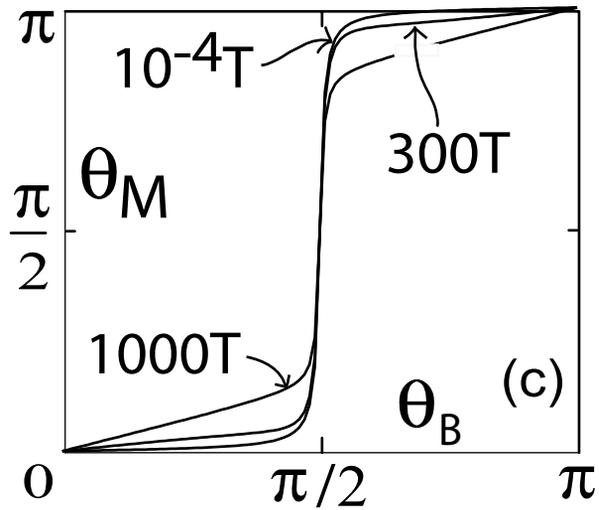

FIG. 1(a). Properties of the p-orbital defect model, Eq. (1). We take crystal-field parameters $V_x = 0$, $V_y = 400K$, $V_z = 2000K$, and spin-orbit coupling $\lambda = -600K$. (a) The six energy levels of the model. The levels do not carry definite angular momentum quantum numbers, but occur in Kramers-degenerate pairs, no matter how strong the spin-orbit coupling. The mixing of the lowest four levels when $|\lambda|$ is comparable to the crystal field parameters $V_{y,z}$ results in a locking of the magnetic moment direction; this locking is not present if $|\lambda| \gg V_{y,z}$ or if $|\lambda| \ll V_{y,z}$. (b) The idea of locking: even if the applied field $B$ is at a large angle $\theta_B$ from the principal axis $z$ of the crystal field, the resultant magnetization vector $M$ lies at a small angle $\theta_M$ from $z$. (c) The calculated $\theta_M$ vs. $\theta_B$ for $|B| = 10^{-4}$ T, 300 T, and 1000 T. For a defect with these parameters, locking is strong for any practical field; it remains strong up to near 1000 T, when the magnetic energy in Eq. (1) becomes comparable to the crystal-field and spin-orbit energies. $M$ unlocks as $\theta_B$ passes through $\frac{\pi}{2}$, rotating rapidly to the opposite direction; however, if $\dot{\theta}_B$ is large enough, this rapid rotation is prevented by Landau-Zener tunneling between the first and second energy levels. (d) The anticrossing of these levels near $\theta_B = \pi/2$; $\frac{E}{|B|}$ is in units of $\mu_B$. The anticrossing gap scales with $|B|$, so that this Landau-Zener tunneling will occur readily at low fields.

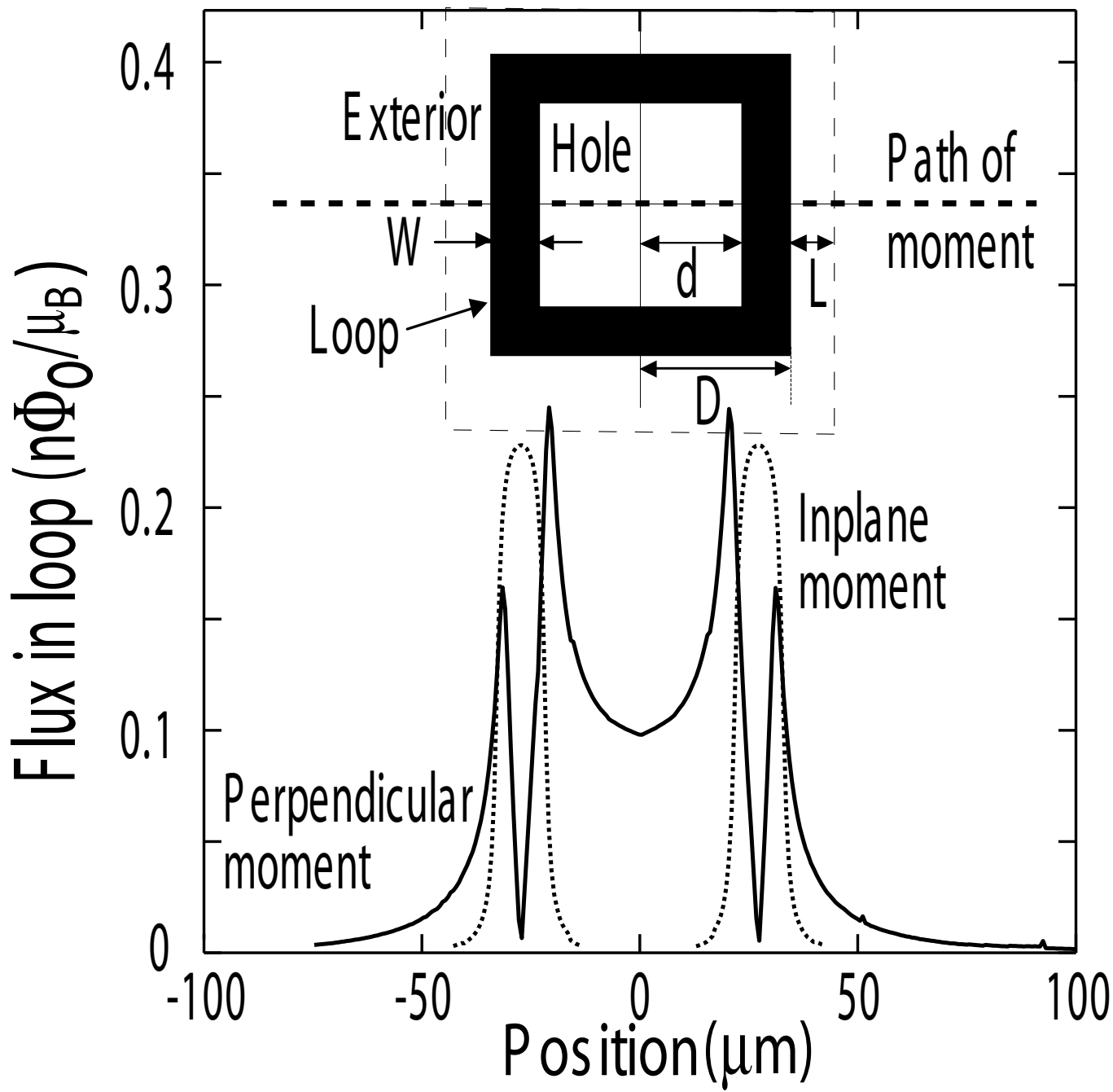

FIG. 2 Magnitude of the flux per Bohr magneton coupled to SQUID loop by a current loop moved along the line indicated. "Inplane" and "perpendicular" refer to the orientation of the magnetic moment. SQUID dimensions are $2D = 52$ μm and $2d = 41.6$ μm. Inset shows configuration of SQUID loop.

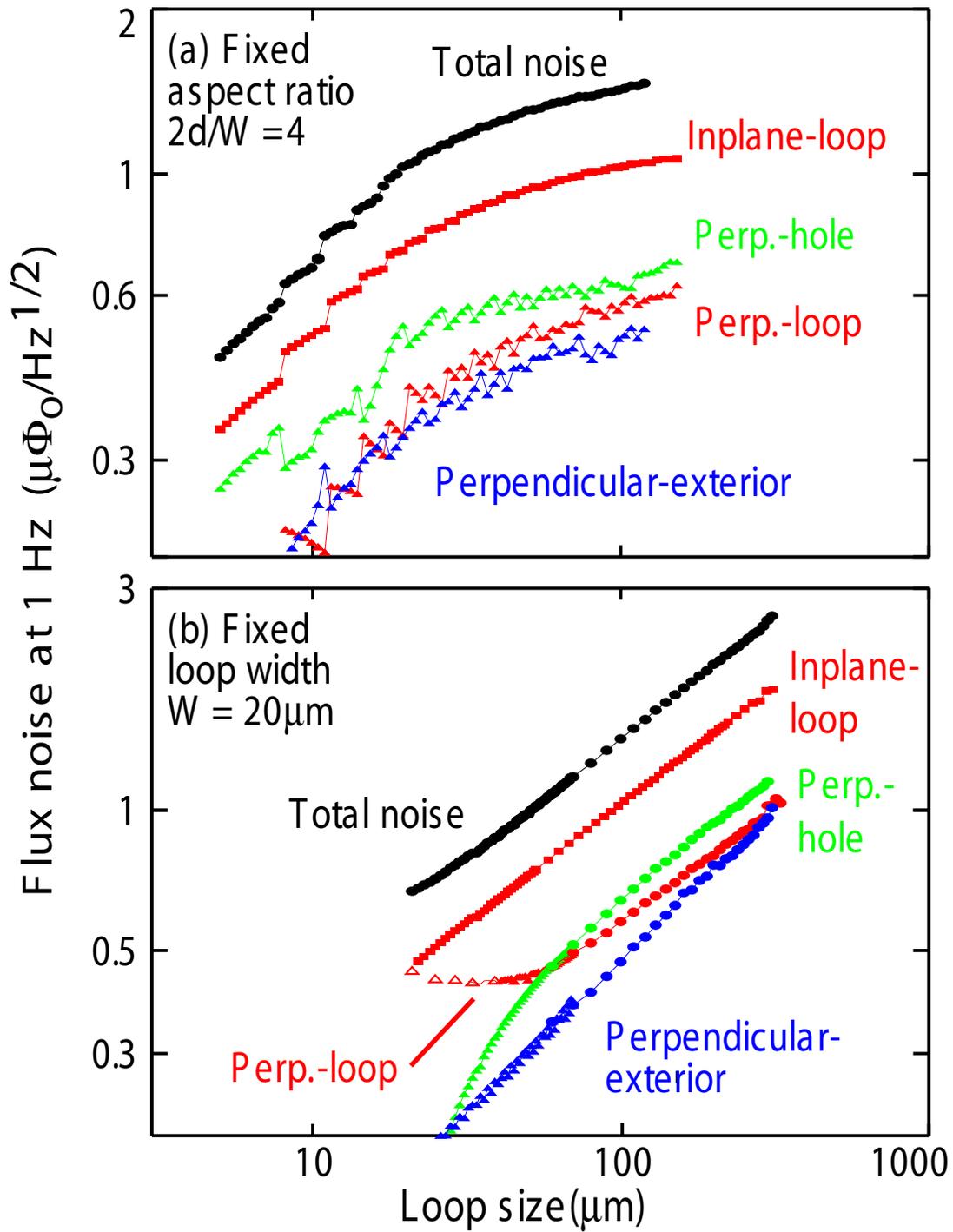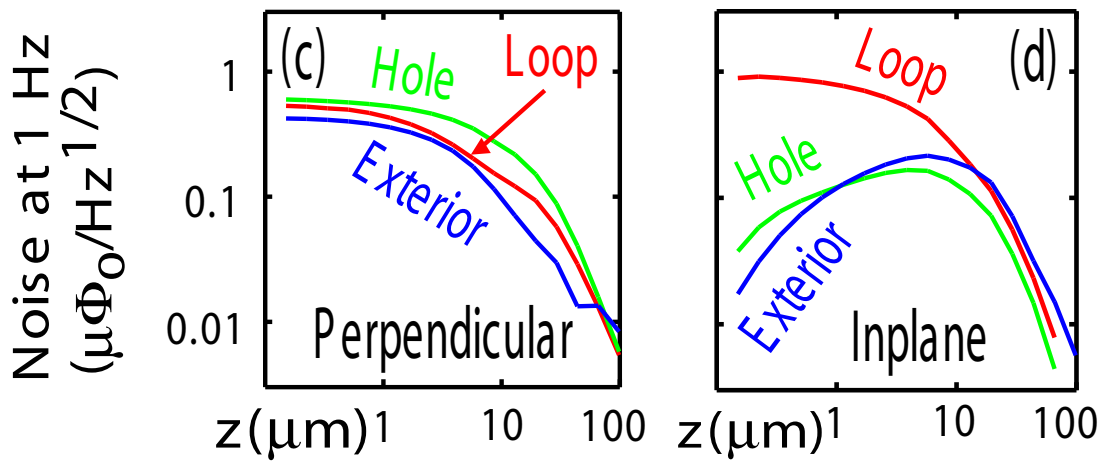

FIG. 3 Computed flux noise versus loop size D + d for (a) fixed loop aspect ratio 2d/W = 4 and (b) fixed width W = 20 μm. The jagged behavior in (a) is due to the discrete mesh of FastHenry. The open triangles in (b) indicate that the accuracy of the calculations is limited.. (c) and (d) show the dependence of the 1/f noise versus the separation of the current and SQUID loops for the perpendicular and one inplane magnetic moment orientation. Dimensions: D = 30 μm, d = 20 μm.